\documentclass[sigconf]{aamas}

\usepackage{graphicx}
\usepackage{subcaption}
\usepackage{amssymb}
\usepackage{hyperref}
\usepackage{multirow}

\setcopyright{ifaamas}  
\acmDOI{}  
\acmISBN{}  
\acmConference[AAMAS'18]{Proc.\@ of the 17th International Conference on Autonomous Agents and Multiagent Systems (AAMAS 2018)}{July 10--15, 2018}{Stockholm, Sweden}{M.~Dastani, G.~Sukthankar, E.~Andr\'{e}, S.~Koenig (eds.)}  
\acmYear{2018}  
\copyrightyear{2018}  
\acmPrice{}  

\begin{document}


\title{Influencing Flock Formation in Low-Density Settings}


\author{Daniel Y. Fu}
\affiliation{
    \institution{Harvard University}
    \city{Cambridge}
    \state{MA}
    \country{USA}
    \postcode{02138}
}
\email{dfu@college.harvard.edu}

\author{Emily S. Wang}
\affiliation{
    \institution{Harvard University}
    \city{Cambridge}
    \state{MA}
    \country{USA}
    \postcode{02138}
}
\email{emilyswang@college.harvard.edu}

\author{Peter M. Krafft}
\affiliation{
    \institution{Massachusetts Institute of Technology}
    \city{Cambridge}
    \state{MA}
    \country{USA}
    \postcode{02139}
}
\email{pkrafft@csail.mit.edu}

\author{Barbara J. Grosz}
\affiliation{
    \institution{Harvard University}
    \city{Cambridge}
    \state{MA}
    \country{USA}
    \postcode{02138}
}
\email{grosz@eecs.harvard.edu}

\begin{abstract}
Flocking is a coordinated collective behavior that results from local sensing
between individual agents that have a tendency to orient towards each other.
Flocking is common among animal groups and might also be useful in robotic
swarms.
In the interest of learning how to control flocking behavior, recent work in
the multiagent systems literature has explored the use of
influencing agents for guiding flocking agents to face a target direction.
The existing work in this domain has focused on simulation settings
of small areas with toroidal shapes.
In such settings, agent density is high, so interactions are common, and
flock formation occurs easily.
In our work, we study new environments with lower agent density, wherein
interactions are more rare.
We study the efficacy of placement strategies and influencing agent behaviors
drawn from the literature, and find that the behaviors that have been shown to
work well in high-density conditions tend to be much less effective in lower
density environments.
The source of this ineffectiveness is that the influencing agents explored in
prior work tended to face directions optimized for maximal influence, but which
actually separate the influencing agents from the flock.
We find that in low-density conditions maintaining a connection to the flock is
more important than rushing to orient towards the desired direction.
We use these insights to propose new influencing agent behaviors, which we dub
``follow-then-influence"; agents act like normal members of the flock to achieve
positions that allow for control and then exert their influence.
This strategy overcomes the difficulties posed by low density environments.
\end{abstract}





\maketitle


\section{Introduction}
\label{sec:introduction}
Flocking behavior can be found in a variety of species across nature, from
flocks of birds to herds of quadrupeds, schools of fish, and swarms of insects.
Researchers have argued that flocking as a collective behavior emerges from
simple, local rules \cite{sumpter2010collective}.
It is therefore natural to imagine placing externally-controlled artificial
agents into flocks to influence them.
Yet it remains an open question whether such techniques are actually effective.
Previous work \cite{genter2013backsearch, genter2013visionstationary, 
genter2014neighborsorientherd, genter2015placement, genter2016facegoalfacecurrent, 
genter201612steplookahead, genterthesis}
has explored the use of influencing agents to guide flocking 
agents to face a target direction in small and toroidal\footnote{In a toroidal
environment, agents that exit the simulation space from one side immediately
re-appear on the other side.} settings, but in such settings, agent density is
high, so interactions are common, and flock formation is rapid.

In the present work, we focus on lower-density settings where interactions are
rarer and flock formation is more difficult.
We study how influencing agent priorities must change in these settings to be
successful and propose new influencing agent strategies to adapt to the
challenges posed by these settings.
Low-density settings are important to study because they capture dynamics in
situations where flocking may not occur naturally, but where we might want to
instigate flocking behavior; imagine a herd of buffalo that is currently
grazing, or a spooked flock of birds where individual agents fail to
coordinate.
Our work may also have implications for coordination in low-density swarms of
robotic multi-agent systems, where control may be imperfect, such as RoboBees
\cite{Chen2017Robobees}.
More broadly, flocking has implications for consensus in animal groups
\cite{Yang2006Consensus, sumpter2008fish, couzin2005} and in human social
networks \cite{liang2012opinion}.
Flocking algorithms have also been used to simulate multivariate timeseries and
human movement \cite{schruben2010multivariate, singham2011agentmovement}.
In all these cases, agent density may vary greatly, so it is important to
understand influencing agent dynamics in both low density and high density
settings.

To study flocking in lower density environments, we introduce two new test
settings.
In one setting, we keep the simulation space toroidal but increase the size of
the space by several factors, greatly decreasing agent density.
Flock formation is still provably guaranteed \cite{jad2003convergence} in this
setting, but is much less rapid, so we study whether influencing agents can
speed up flock formation.
In the second setting, similar to existing ``sheep herding'' tasks
\cite{nalepka2017herd}, we use a non-toroidal simulation space and start the
flocking agents inside a circle in the center.
Since this space is non-toroidal, flock formation is not guaranteed, so we
study whether influencing agents can instigate flocking behavior by keeping the
flocking agents in a pre-defined area, or by moving them all in a certain
direction.

We find that results from the existing literature are not robust in these
environments with low agent density, since agent interactions are more rare.
In particular, with less frequent local interactions between agents,
maintaining a connection to the flock becomes a key factor in the efficacy of
influencing agent behaviors.
Simple behaviors such as ``face the goal direction" are often superior to more
complex behaviors that try to optimize for speed of convergence.
We experiment with a number of new strategies and find that a multi-stage
approach of ``follow-then-influence" is most effective in low-density
environments.
In this approach, influencing agents start out by participating as normal
members of the group, embedding themselves inside small, naturally-forming
flocks.
After some time, the influencing agents start pushing their neighbors to face a
given goal direction.

The main contributions of this work are:
\begin{itemize}
    \item An investigation of two new low-density flocking settings, where
    flock formation is more difficult.
    \item The introduction of new influencing agent behaviors to adapt to the
    difficulties presented by these new settings.
    \item Analysis of the major differences in influencing agent priorities in
    low-density vs. high-density settings.
\end{itemize}

The rest of this paper is organized as follows: \S\ref{sec:problem} describes 
the flocking model we use and the new test settings.
\S\ref{sec:influencing} describes the role of influencing agents and formalizes
agent behaviors.
\S\ref{sec:experimental} describes our experimental setup and the experiments 
we run to evaluate agent behaviors in the new test settings, and
\S\ref{sec:evaluation} presents the results.
Finally, we conclude and discuss future work in \S\ref{sec:conclusion}.

\graphicspath{{images/}}

\section{Problem Description}
\label{sec:problem}
\subsection{Flocking Model}
\label{sec:model}
Like other studies in the literature, we use a simplified version of Reynold's 
Boid algorithm \cite{reynoldsmodel} to model the flock.
In this simplified model, also proposed independently by Vicsek and
collaborators \cite{vicsek1995}, 
agents change their alignment at every step to be similar to the average 
alignment of other agents in their neighborhood.
At each time step, each agent $i$ moves with constant speed $s=0.7$ and has 
orientation $\theta_i(t)$ with position $p_i(t) = (x_i(t), y_i(t))$.
At timestep $t$, agent $i$ updates its position based on its alignment: 
$x_i(t) = x_i(t-1) + s\cos(\theta_i(t))$ and 
$y_i(t) = y_i(t-1) + s\sin(\theta_i(t))$.
At the same time, the agents change their orientation based on the alignments 
of neighboring agents.
Let the neighbors $N_i(t)$ be the set of agents at time $t$ that are within 
neighborhood radius $r$ of agent $i$, not including agent $i$ itself.
At timestep $t$, each agent updates its orientation to turn towards the average
of its neighbors' orientations:
\[\theta_i(t+1)=\theta_i(t)+\frac{1}{2}\frac{1}{|N_i(t)|} \Sigma_{j \in N_i(t)}
\left(\theta_j(t) - \theta_i(t)\right)\]
The factor of $\frac{1}{2}$ in the second term reflects a ``momentum" factor.

\subsection{New Settings}
\label{sec:settings}
Previous work has studied influencing agents in a small toroidal 150 $\times$ 150
grid, with neighborhood radius $r=20$~\cite{genter2016facegoalfacecurrent,
genter201612steplookahead}. 
In this work, we study low-density dynamics by introducing two new settings
that are more adverse to flock formation; we call these new settings the
\textit{large} setting and the \textit{herd} setting.
In these settings, we set the neighborhood radius to $r=10$.

In the \textit{large} setting, non-influencing agents are randomly placed in a 
toroidal 1,000 $\times$ 1,000 grid with random initial orientations.
The larger grid size results in lower agent density; as a result, agents start
out much farther away from other agents' neighborhoods, and interactions are
much rarer.
However, since the simulation space remains toroidal, convergence to a single
direction is still provably guaranteed, so we are primarily interested in
studying the length of time to convergence in this case
\cite{jad2003convergence}.
In the \textit{herd} setting, non-influencing agents are placed randomly in a 
circle of radius 500 whose origin lies at the center of a 5,000 $\times$ 5,000
non-toroidal grid.
When agents reach the edge of the grid, they move off-world; in this way,
agents can get ``lost" from the rest of the flock.
As a result, convergence to a single flock is not guaranteed.
Therefore, we are interested in studying how well influencing agents can keep
the non-influencing agents from getting lost.

\section{Influencing Agents}
\label{sec:influencing}
We can change flock dynamics by introducing influencing agents that we control.
We refer to non-influencing agents as Reynolds-Vicsek agents.
We do not give the influencing agents any special control over the
Reynolds-Vicsek agents; we simply let them interact with influencing
agents using the same local sensing rules as with any other agent.
We also limit the influencing agents to have the same speed as Reynolds-Vicsek
agents, both to help the influencing agents ``blend in" in real applications
and to be consistent with the related prior work.
We let the influencing agents have a sensing radius of twice the normal
neighborhood radius.
This allows the influencing agents to see their neighbors' neighbors,
allowing for more complex algorithms.
In some cases, the influencing agents can communicate with each other, but
they do not need a global view.

\subsection{Placement}
\label{sec:placement}
\begin{figure}
    \centering
    \includegraphics[width=0.5\textwidth]{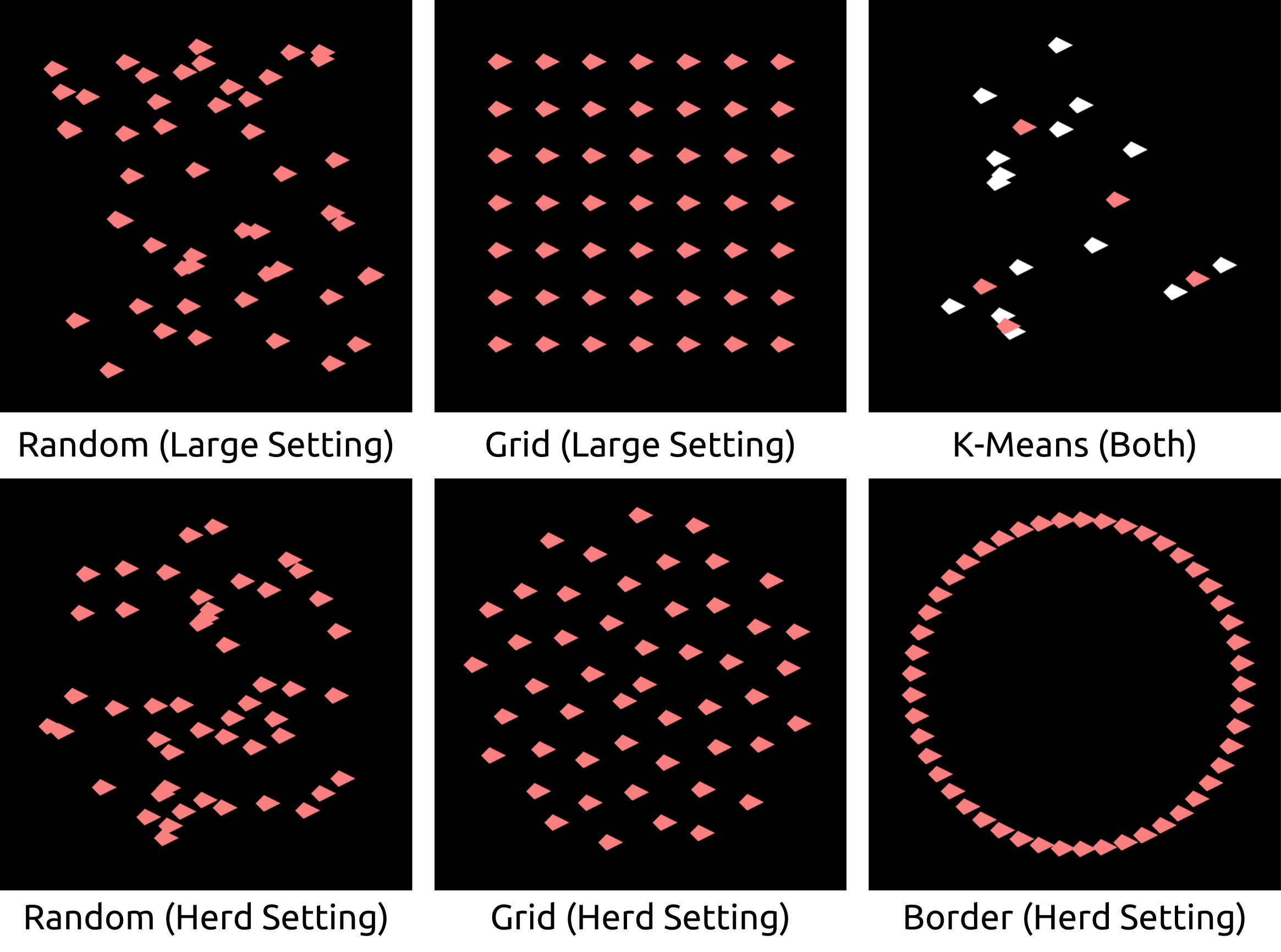}
    \caption{The different placement strategies we explore in this paper.
    Red agents are influencing agents, and white agents are Reynolds-Vicsek
    agents.
    Note that \textit{k-means} is the only placement strategy where the
    placement of influencing agents depends on placement of Reynolds-Vicsek
    agents.}
    \label{fig:placements}
\end{figure}

Each influencing agent algorithm we use is decomposed into two parts:
a placement strategy and an agent behavior.
Except for slight modifications to make some of these strategies work in a
circular environment, the placement strategies we use are drawn from the
literature \cite{genter2015placement, genterthesis}.
The placement strategies we use in this work are shown in Figure
$\ref{fig:placements}$.
We note that the question of how to maneuver influencing agents to reach the
positions given by these placement strategies is important, but out of scope
for this paper.
For a discussion of this question, we refer the reader to Genter and Stone 
\cite{genter2016facegoalfacecurrent, genterthesis}.

We use three placement strategies for the \textit{large} setting:
\textit{random}, \textit{grid}, and \textit{k-means}.
The \textit{random} placement strategy, as its name suggests, places
influencing agents randomly throughout the grid.
The \textit{grid} placement strategy computes a square lattice on the grid and
places influencing agents on the lattice points.
This strategy ensures regular placement of influencing agents throughout the
grid.
The \textit{k-means} placement strategy uses a $k$-means clustering algorithm
on the positions of Reynolds-Vicsek agents in the simulation space.
This strategy finds a cluster for each influencing agent by setting $k$ equal
to the number of influencing agents, and then places an influencing agent at
the center of each cluster.

We develop similar placement strategies for the \textit{herd} setting, with
some slight modifications.
To adapt the strategies to a circular arrangement of agents, we define each
strategy in terms of some radius $r$ about an origin $O$, except for the
\textit{k-means} strategy, which remains the same.
We modify the \textit{random} placement strategy to randomly distribute agents
within the circle of radius $r$ about the origin $O$, instead of the entire
simulation space.
We adapt the \textit{grid} placement strategy to a circular setting using a
sunflower spiral \cite{segermansunflower}.
In polar coordinates relative to $O$, the position of the $n$-th influencing
agent in a sunflower spiral is given by $(c\sqrt{n}, \frac{2\pi}{\phi^2}n)$,
where $\phi$ is the golden ratio, and $c$ is a normalizing constant such that
the last influencing agent has distance $r$ from $O$.
We also introduce a circular placement strategy, inspired from the border
strategies used in prior work \cite{genter2015placement}.
This strategy places agents on the circumference of the circle of radius $r$
around the origin $O$.
We refer to the circular strategies as \textit{circle-random},
\textit{circle-grid}, and \textit{circle-border}, respectively.

\subsection{Behaviors}
\label{sec:behaviors}
Once we have placed the influencing agents, we still need to design how they
will work together to influence the flock.
We call this aspect of the design ``agent behaviors."
In the present work we focus on decentralized ``ad-hoc" algorithms for the
influencing agents, since this class of algorithms has been the focus of the
existing multiagent systems literature on this topic
\cite{genterthesis, genter2014neighborsorientherd, genter201612steplookahead}.
A summary of the behaviors we investigate is shown in Table
\ref{table:behaviors}.

\begin{table}[]
\footnotesize
\centering
\caption{Summary of the behaviors we investigate}
\label{table:behaviors}
\begin{tabular}{|l|l|l|l|}
\hline
\textbf{Setting}       & \textbf{Goal Type}               & \textbf{Name}        & \textbf{Description}                    \\ \hline
\multirow{5}{*}{Large} & \multirow{3}{*}{\textbf{}}  & \textit{Face}      & Always face goal direction        \\ \cline{3-4}
                       &                             & \textit{Offset Momentum}      & Offset last average velocity      \\ \cline{3-4}
                       &                             & \textit{One-Step Lookahead}  & Simulate one step, choose best      \\ \cline{3-4}
                       &                             & \textit{Coordinated}      & Pair off and coordinate      \\ \cline{3-4}
                       &                             & \textit{Multistep}   & \textit{\textbf{Follow-then-influence}} \\ \hline
        \multirow{7}{*}{Herd}  & \multirow{4}{*}{Traveling} & \textit{Face}      & (As above)        \\ \cline{3-4}
                       &                             & \textit{Offset Momentum}      & (As above)      \\ \cline{3-4}
                       &                             & \textit{One-Step Lookahead}  & (As above)      \\ \cline{3-4}
                       &                             & \textit{Coordinated}      & (As above)      \\ \cline{2-4}
                       & \multirow{3}{*}{Stationary} & \textit{Circle}      & Trace circle around agents              \\ \cline{3-4}
                       &                             & \textit{Polygon}     & Trace polygon around agents             \\ \cline{3-4}
                       &                             & \textit{Multicircle} & \textit{\textbf{Follow-then-influence}} \\ \hline
\end{tabular}
\end{table}

\subsection{Large Setting}

For the \textit{large} setting, we study four behaviors drawn from prior work
\cite{genter201612steplookahead,genter2016facegoalfacecurrent,
genter2015placement}, and one new \textit{multistep} behavior.

In previous work, Genter and Stone have introduced baselines \textit{face} and
\textit{offset momentum} behaviors, as well as more sophisticated
\textit{one-step lookahead} and \textit{coordinated} behaviors.
Each of these behaviors aims to turn Reynolds-Vicsek agents to a pre-set goal
angle $\theta^*$
Influencing agents using the \textit{face} behavior always face the angle
$\theta^*$.
With the \textit{offset momentum} behavior, influencing agents calculate the
average velocity vector of the agents in their neighborhood, and align to a
velocity vector that, when added to the average velocity vector, sums to the
vector pointing in direction $\theta^*$.
We note that such a vector always exists; if the average velocity vector is
$(x, y)$, and $\theta^*$ is represented by vector $(x', y')$, then the agents
align to vector $(x'-x, y'-y)$.
A \textit{one-step lookahead} influencing agent cycles through different
angles and simulates one step of each of its neighbors if it were to move in
that angle.
It adopts the angle that results in the smallest average difference in angle
from $\theta^*$ among all its neighbors.
Finally, with the \textit{coordinated} behavior, each agent pairs with another
and runs a one-step lookahead to minimize the average difference in angle from
$\theta^*$ among both their neighbors.
For a more detailed explanation of these behaviors, especially the
\textit{coordinated} behavior, we direct the reader to Genter and Stone
\cite{genter201612steplookahead}.

The \textit{multistep} behavior is a novel contribution and adopts
what we call a ``follow-then-influence" behavior.
In the initial stage, influencing agents simply behave like normal Reynolds-Vicsek
agents; as a result, they easily join flocks and become distributed throughout
the grid.
At the same time, each influencing agent estimates how many Reynolds-Vicsek agents
are path-connected to it; here, we define two agents as being path-connected if
there is a path between them, where edges are created by two agents being in
each other's neighborhood.
An accurate calculation of path-connectedness requires a global view
from every influencing agent, since paths may extend arbitrarily far away from
the influencing agent.
In our algorithm, we only consider Reynolds-Vicsek agents that are within the
sensing radius of the influencing agents.
Given their local estimates, the influencing agents compute a global sum of all
their estimates; once that sum passes over some threshold $T$, the influencing
agents calculate the average angle $\overline{\theta}$ among all the agents that
are locally connected to influencing agents, and from there adopt the
\textit{face} behavior with goal direction $\overline{\theta}$.

We also explore some variations of the \textit{multistep} behavior by noticing
that once the sum of connected agents passes the threshold $T$, any of the
other behaviors studied can be used to turn the Reynolds-Vicsek agents towards
the final goal direction $\overline{\theta}$.
In other words, the \textit{multistep} behavior can be paired with any other
behavior.
We study these pairings to see how effective they are.

\subsection{Herd Setting}
For the \textit{herd} setting, we divide the behaviors into two categories:
traveling behaviors and stationary behaviors.
As a reminder, in the \textit{herd} setting, the simulation space is
non-toroidal, and all the Reynolds-Vicsek agents start in a circle in the
center.
In this setting, flock formation is not guaranteed, so we are interested in
using influencing agents to instigate flocking behavior.
There are two different choices we can make; we can either try to force the
Reynolds-Vicsek agents to stay in the center (stationary behaviors),
or we can let the influencing agents direct the Reynolds-Vicsek agents away
from their initial starting position (traveling behaviors).
Since all the agents have a constant speed, the former is much more difficult
than the latter, so we must evaluate them separately.

For the traveling behaviors, we can use all the behaviors used in the
\textit{large} setting, except for the \textit{multistep} behavior.
Since the world is non-toroidal, it is not guaranteed that the number of
connected agents will ever pass the threshold $T$; in this case, the
influencing agents would simply wander forever.

We study three stationary behaviors: \textit{circle}, \textit{polygon},
and \textit{multicircle}.
The \textit{circle} and \textit{polygon} behaviors have each influencing agent
trace a circle or polygon around the origin.
For placement strategies where influencing agents have different distances to
the origin, the influencing agents simply trace circles and polygons of
different radii.

The \textit{multicircle} behavior is analogous to the \textit{multistep}
behavior from \textit{large}.
The influencing agents start out by circling around the origin and wait for
Reynolds-Vicsek agents to enter their neighborhood.
Once they detect Reynolds-Vicsek agents in their neighborhood, they adopt a
``following" behavior where they act like Reynolds-Vicsek agents to integrate
into a small flock.
They continue this following stage until reaching a final radius $r_F$, at which
point they again adopt a circling behavior.
In addition to building influence by following before influencing, this behavior
also makes maintaining influence easier; since the final radius is larger than
the original radius, the final path turns less sharply than if the influencing
agents had stayed at their original radius.
To the best of our knowledge, this is the first presentation of such a
multi-stage behavior to induce circling behavior under the Reynolds-Vicsek
model in the literature.

\section{Experimental Setup}
\label{sec:experimental}
We extended the MASON simulator to run the experiments \cite{luke05mason}.
We used the default parameters for the Flocking simulation that is included
with the MASON simulator, except without any randomness, cohesion, avoidance,
or dead agents.
We sampled all metrics every 100 time steps and ran all experiments for 100
trials.

\subsection{No Influencing Agents}
Previous literature compared new influencing agent behaviors with baseline 
influencing agent behaviors, but did not compare to settings with no
influencing agents.
In order to observe the marginal contribution of influencing agents 
in future experiments, we start our investigation of the \textit{large} and
\textit{herd} settings by studying flock formation in those environments without
any influencing agents.
We use two metrics to understand flock formation: average number of flocks 
formed and average proportion of lone agents at each time step.

In the \textit{large} setting, we test on a 1,000 $\times$ 1,000 grid and vary the
number $N$ of Reynolds-Vicsek agents from 50 to 300 in increments of 50.
We run these simulations for 6,000 time steps.
In the \textit{herd} setting, we use a 5,000 $\times$ 5,000 grid, position the
herd in the center of the grid with radius 500, and vary $N$ from 50 to 300 in
increments of 50. 
We run these simulations for 6,000 time steps.

\subsection{Influencing Agents}
To evaluate the contributions of influencing agents in the \textit{large}
setting, we measure time to convergence.
We define convergence as having half the Reynolds-Vicsek agents face the same
direction, since full convergence takes much longer.

We test the \textit{random}, \textit{grid}, and \textit{k-means} placement
strategies, along with the full suite of behaviors in the \textit{large}
setting.
We place 300 Reynolds-Vicsek agents on the grid and vary
the number of influencing agents from 10 to 100 in intervals of 10.

To evaluate the contributions of influencing agents in the \textit{herd}
setting, we measure a slightly different metric.
Since we have two qualitatively different categories of behaviors (traveling
behaviors vs. stationary behaviors), the number of agents facing the same
direction is irrelevant.
The stationary behaviors rotate the agents around
the origin (in fact, if the Reynolds-Vicsek agents are all facing the same
goal direction, the stationary behavior has failed).
Instead, we measure the number of Reynolds-Vicsek agents that are connected to
influencing agents at 15,000 time steps; at this point in time, all the agents
have travelled out of the grid, and no new interactions occur.
As a result, this quantity measures sustained influence over the
Reynolds-Vicsek agents over time.

In the \textit{herd} setting, we examine the three circular placement
strategies---\textit{circle-border}, \textit{circle-random}, and
\textit{circle-grid}---with two placement radii, 500 and 750,
along with the \textit{k-means} placement strategy.
We split our examination of behaviors between the traveling behaviors
(the same behaviors as used in the \textit{large} setting, minus the
\textit{multistep} behavior) and three stationary behaviors---\textit{circle},
\textit{polygon}, and \textit{multicircle}.
We use a polygon with ten sides (a decagon) for the \textit{polygon}
behavior, and we vary the final radius for the \textit{multicircle} behavior
based on the initial placement radius.
When the placement radius is 500, we set the final radius to
900; when the placement radius is 750, we set the final
radius to 1,100.
We place 300 Reynolds-Vicsek agents on the grid and again vary the number of
influencing agents from 10 to 100 in intervals of 10.

\section{Results}
\label{sec:evaluation}

\subsection{No Influencing Agents}
\begin{figure*}
    \centering
    \includegraphics[height=0.19\textwidth]{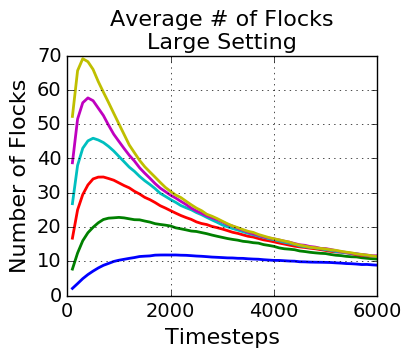}
    \includegraphics[height=0.19\textwidth]{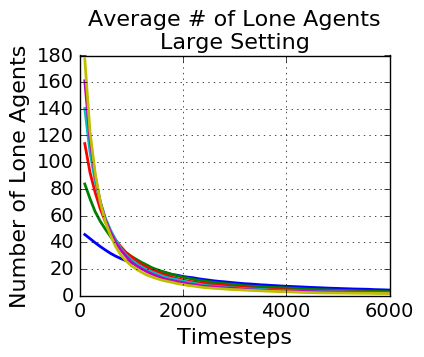}
    \includegraphics[height=0.19\textwidth]{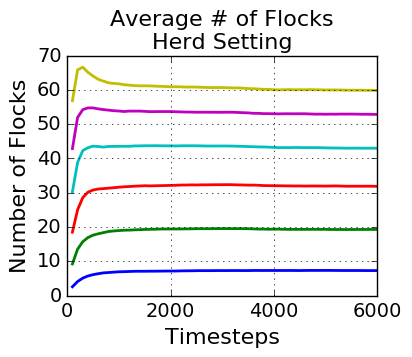}
    \includegraphics[height=0.19\textwidth]{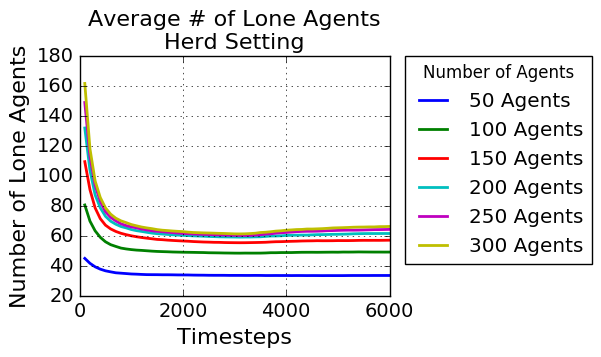}
    \caption{Average flock counts and lone agent counts over time for the
    \textit{large} and \textit{herd} settings with no influencing agents,
    varying the number of Reynolds-Vicsek agents.}
    \label{fig:no_adhoc}
\end{figure*}

First, we briefly characterize the flocking behavior of a group of Reynolds-Vicsek
agents without influencing agents in the \textit{large} and \textit{herd} settings.
We measure the number of clusters of agents that are path-connected and facing the
same direction; each of these clusters forms a small flock.
We also measure the number of lone agents (the number of agents with no neighbors).
Figure $\ref{fig:no_adhoc}$ shows graphs of these values over time for the two
settings.

In the \textit{large} setting, there are two qualitative stages of convergence:
initial flock formation and flock unification.
In the first stage, individual agents collide with each other and form small 
flocks, so the number of flocks increases.
In the second stage, these small flocks that formed collide with one another and
join together to form larger flocks, so the number of flocks decreases.
In Figure $\ref{fig:no_adhoc}$, the first stage is represented by the initial
increase in the average number of flocks, and the second stage is represented
by the following decrease in the average number of flocks.
This behavior is reflected in the continually decreasing number of lone agents;
since the number of lone agents continues to decrease over time, we know that the
decrease in the total number of flocks is due to flock convergence.
Note that when there are more total agents, the absolute number of lone agents
decreases faster and reaches a similar value to the other cases by the end of the
simulation.
In other words, the \textit{ratio} of lone agents to total agents hits a 
lower value when there are more agents, but the final absolute number of total 
lone agents is still similar to the other cases.

The two stages of convergence also occur somewhat in the \textit{herd} setting,
but the second stage is cut off by the non-toroidal nature of the setup.
As flocks leave the starting area, the chances of interacting with other flocks 
vastly decreases, so most of the flocks formed from the first stage never end up
merging with other flocks.
This is reflected in the plateaus of both the total number of flocks and the total
number of lone agents.
One small artifact in the metric is worth mentioning; since the agents start off
in a much smaller area than in the \textit{large} setting, many of the agents
start out with a non-zero number of neighbors.
This causes the initial value of the average number of flocks to be non-zero, and
the average number of lone agents to be less than the total number of agents.

\subsection{Influencing Agents in the Large Setting}
\begin{figure*}
    \centering
    \begin{subfigure}[b]{0.7\textwidth}
        \includegraphics[width=\textwidth]{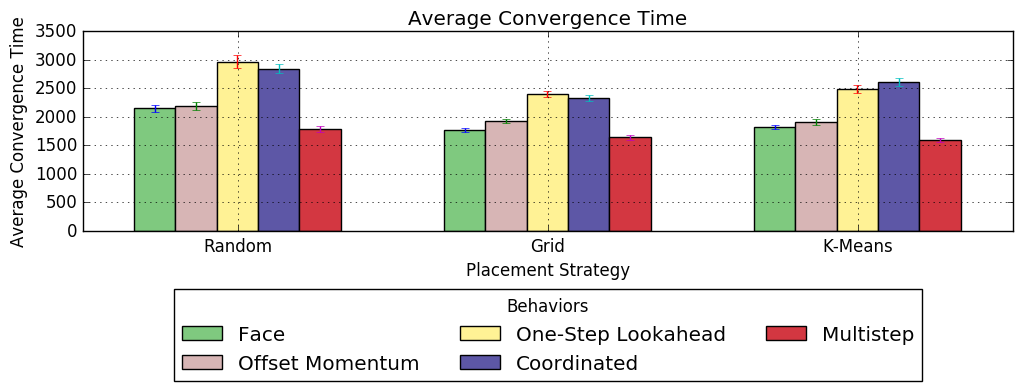}
    \end{subfigure}
    \begin{subfigure}[b]{0.7\textwidth}
        \includegraphics[width=\textwidth]{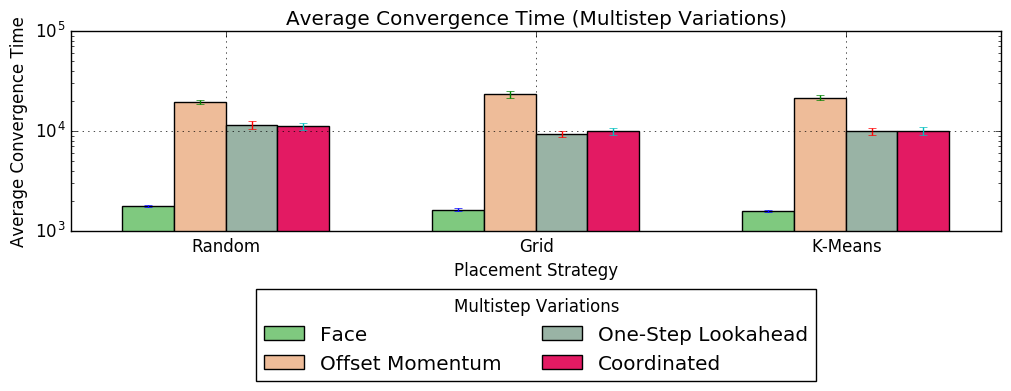}
    \end{subfigure}
    \caption{Average times to 50\% convergence for 300 Reynolds-Vicsek agents
    with 50 influencing agents in the \textit{large} setting under different
    placement strategies and behaviors.
    Top: The five main behaviors for the \textit{large} setting.
    Bottom: Variations on the \textit{multistep} behavior, in log scale.
    Smaller is better.
    Error bars show standard error of the mean.}
    \label{fig:large}
\end{figure*}

Next, we report on the efficacy of the behaviors in the \textit{large} setting.
The average times for $50\%$ convergence with different placement strategies
and the five behaviors are shown in Figure $\ref{fig:large}$.
We show graphs for 300 Reynolds-Vicsek agents and 50 influencing agents only,
since the trends for the other numbers of influencing agents were similar (the
major difference being that when there are more influencing agents, convergence
happens faster, and when there are fewer influencing agents, convergence
happens slower).
Note that smaller is better in these graphs.

The most immediately striking finding is that, in less dense settings,
the \textit{one-step lookahead} and \textit{coordinated} behaviors
significantly underperform the ``baseline" \textit{face} and
\textit{offset momentum} behaviors, irrespective of placement strategy.
This is an opposite result from Genter and Stone's findings on smaller simulation
spaces \cite{genter201612steplookahead, genterthesis}, which found that the
\textit{one-step lookahead} and \textit{coordinated} behaviors outperform the
\textit{face} and \textit{offset momentum} behaviors.
This finding is also rather counterintuitive; why should the ``smarter" behaviors
underperform the simpler behaviors?

The answer is that, when agent interactions are rare, it is more important for
influencing agents to \textit{maintain influence} than it is for them to quickly
change the direction of neighboring Reynolds-Vicsek agents.
The \textit{one-step lookahead} and \textit{coordinated} behaviors underperform
here because they tend to send influencing agents away from neighboring agents.
\begin{figure}
    \includegraphics[width=0.48\textwidth]{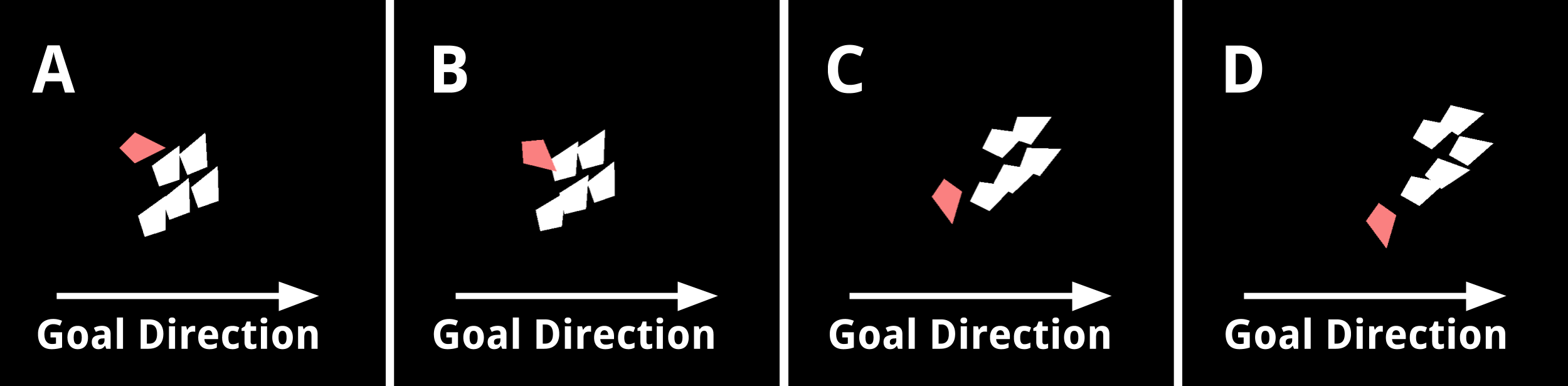}
    \caption{An example of an influencing agent losing influence under the
    \textit{one-step lookahead} behavior.
    The influencing agent is shown in red, and the Reynolds-Vicsek agents are
    shown in white.
    In A, the influencing agent first encounters the flock of Reynolds-Vicsek
    agents.
    In B-D, the influencing agent takes on directions that are oriented away
    from the goal direction to try to rapidly influence the Reynolds-Vicsek
    agents.
    This changes the orientation of the Reynolds-Vicsek agents, but the
    influencing agent has started to travel away from the flock by D.}
    \label{fig:losing_influence}
\end{figure}
An example of this phenomenon is shown in Figure $\ref{fig:losing_influence}$.
The influencing agent, shown in red, adopts an orientation that turns
neighboring Reynolds-Vicsek agents towards the goal direction.
Even though this action does turn Reynolds-Vicsek agents towards the goal
direction, the influencer cannot successfully turn all the agents in a single
step; as a result, the influencing agent must maintain that orientation for
future steps.
However, as long as the neighboring agents are not facing the goal direction,
the influencing agent's chosen orientation takes it away from the center of the
flock of Reynolds-Vicsek agents, causing the agent to lose influence.
Once the influencing agent has lost influence, the agent has difficulty
catching up with the same flock, since influencing agents travel at the same
speed\footnote{There are some approaches which remove this speed constraint
from influencing agents \cite{han2010teleporting}.
However, this choice allows for unrealistic behaviors wherein influencing
agents travel to one Reynolds-Vicsek agent at a time and change the direction
of the individual Reynolds-Vicsek agent before moving on to the next one.
This behavior results in Reynolds-Vicsek agents that are all facing the same
direction, but that are often not path-connected.}
as Reynolds-Vicsek agents.
As a result, the influencing agent is not actively influencing the direction of
any Reynolds-Vicsek agents until it encounters another group of Reynolds-Vicsek
agents.

Note that this effect also happens on a smaller simulation space, but it is
not nearly as pronounced; when interactions are very frequent, influencing
agents that have lost influence can find another group of Reynolds-Vicsek
agents very quickly.
As a result, the gains from the smarter local algorithm still outweigh the
negative effects from losing influence.

The \textit{multistep} behavior does not suffer from the same problem; it
can both maintain influence and effectively turn Reynolds-Vicsek agents and so
outperforms all the other behaviors by a couple hundred steps.
When the \textit{multistep} behavior is paired with the other behaviors,
though, it magnifies their inability to maintain influence.
The bottom graph of Figure \ref{fig:large} shows variations on the
\textit{multistep} behavior, wherein influencing agents adopt different
behaviors after the number of Reynolds-Vicsek agents under control passes $T$.
Note that the variations that pair the \textit{multistep} behavior with the
\textit{offset momentum}, \textit{one-step lookahead}, and \textit{coordinated}
behaviors perform almost an order of magnitude worse than the
\textit{multistep-face} behavior.
What is the root cause of this difference?
The \textit{multistep} behavior starts out by creating many local flocks, some
of which have influencing agents in them.
When interactions are rare, the \textit{offset momentum}, \textit{one-step
lookahead}, and \textit{coordinated} behaviors have difficulty changing the
orientation of existing flocks quickly before losing influence.
As a result, the \textit{multistep} behavior takes an order of magnitude longer
to reach convergence when paired with the other behaviors.

Finally, we note that the effect of placement behaviors on convergence time
are almost non-existent.
When the density is lower, there is a much smaller chance that any influencing
agent will start out with more than one Reynolds-Vicsek agent in its
neighborhood, even with the \textit{k-means} placement behavior.
As a result, even the best clustering approach is almost the same as starting
out randomly or in a grid.

\subsection{Influencing Agents in the Herd Setting}
\begin{figure*}
    \centering
    \begin{subfigure}[b]{0.7\textwidth}
        \includegraphics[width=\textwidth]{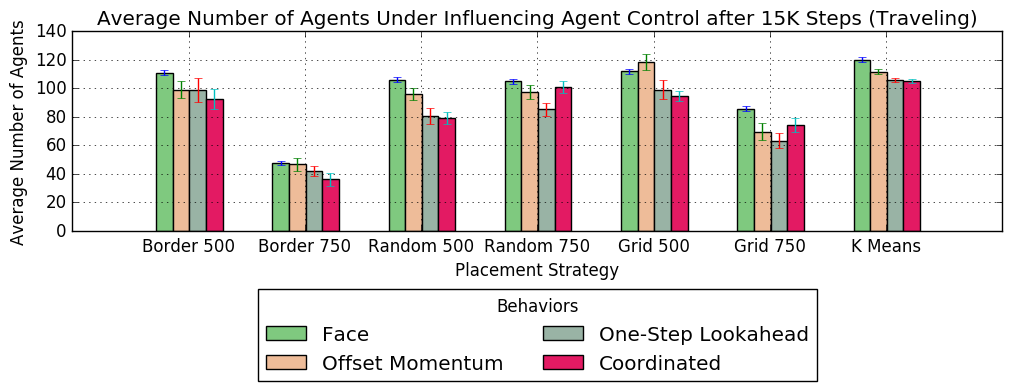}
    \end{subfigure}
    \begin{subfigure}[b]{0.7\textwidth}
        \includegraphics[width=\textwidth]{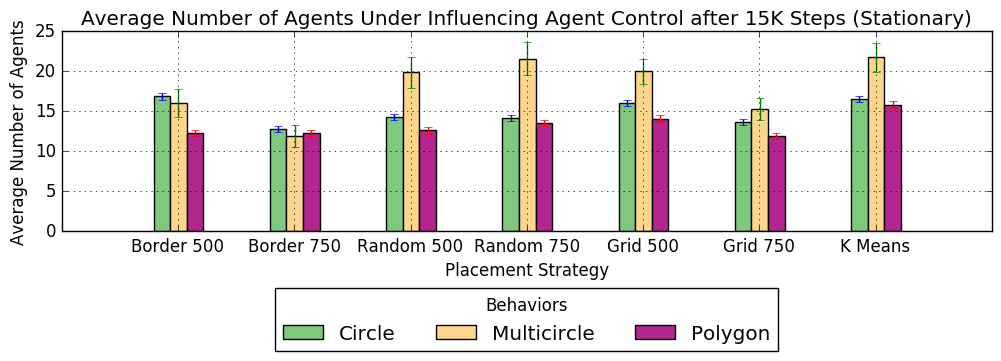}
    \end{subfigure}
    \caption{Average number of agents under influencing agent control after 15,000
    steps with 300 Reynolds-Vicsek agents and 50 influencing agents in the
    \textit{herd} setting under various placement strategies
    and influencing agent behaviors.
    The traveling behaviors attempt to control the direction of the Reynolds-Vicsek
    agents, while the stationary behaviors keep the influencing agents near
    the goal area using circling techniques.
    Larger is better.
    Error bars represent standard error of the mean.}
    \label{fig:herd}
\end{figure*}

Next, we evaluate results for our experiments in the \textit{herd} setting.
In many cases, measuring the number of agents facing the same direction is not
interesting here, since it is impossible to keep Reynolds-Vicsek agents in one place
if they are facing the same direction.
Instead, we exclusively measure the number of Reynolds-Vicsek agents that are
path-connected to influencing agents and facing the same direction as the influencing
agent.
This is a measure of ``control" of the Reynolds-Vicsek agents.
The average number of agents in such local flocks after 15,000 time steps is given
in Figure $\ref{fig:herd}$ for both the traveling and stationary behaviors.
We find that the traveling behaviors vastly outperform any of the stationary
behaviors.
However, there may be environments in reality for which the traveling behaviors
are not applicable (suppose it is strictly necessary to keep a flock in one
place, for instance).
Thus, we analyze the traveling behaviors separately from the stationary behaviors.

\subsubsection{Traveling}
Again, we find that the \textit{face} behavior tends to outperform the
\textit{offset momentum}, \textit{one-step lookahead}, and \textit{coordinated}
behaviors; we attribute this to the tendency of the \textit{offset momentum},
\textit{one-step lookahead}, and \textit{coordinated} behaviors to lose
influence over time.
We note that the effect is not as pronounced here as in the \textit{large}
experiments, since each influencing agent has to control fewer agents.

In contrast to the \textit{large} experiments, we find that here the placement
strategy has a major impact on the efficacy of the traveling behaviors.
Again, this has to do with density of influencing agents.
For example, notice that \textit{Border 750} (place the influencing agents in a
circle about the origin with radius 750) vastly underperforms the other
placement strategies.
The larger radius results in a lower density of influencing agents, so a greater
number of Reynolds-Vicsek agents slip through the ``holes."
Furthermore, by the time the Reynolds-Vicsek agents reach the border, they have
already formed flocks, and it is more difficult for the influencing agents to
point them in the right direction.
This effect is less pronounced for \textit{Grid 750} and almost non-existant
for \textit{Random 750}, since these strategies place influencing agents within
the circle, and not simply along its circumference.
As a result, the Reynolds-Vicsek agents still encounter influencing agents
before reaching the circumference of the circle.

Finally, we note that \textit{k-means} outperforms all other placement strategies
by a few agents.
Again, the main driving factor behind this is agent density.
When an influencing agent starts out in a clustered area, it has at least one
other Reynolds-Vicsek agent in its neighborhood.
As a result, its effective area of influence is slightly larger than with the
other placement strategies.
This helps it pick up more Reynolds-Vicsek agents.

\subsubsection{Stationary}
Among the stationary behaviors, the \textit{multicircle} behavior achieves the
best outcomes, but its performance depends on its paired placement strategy.
The \textit{multicircle} behavior slightly underperforms the \textit{circle}
behavior when paired with the \textit{Border} placement strategies; slightly
overperforms when paired with the \textit{k-means}, \textit{Random}, and
\textit{Grid 500} placement strategies; and performs the same as
\textit{circle} in the \textit{Grid 750} strategy.
What drives these trends?
Once the \textit{multicircle} behavior reaches the final stage, it is tracing a
larger circle than the \textit{circle} behavior traces on its own.
As a result, it is easier to maintain influence and turn the Reynolds-Vicsek
agents over time in the final stage.
Before that, however, the influencing agents are in a following stage.
When the influencing agents start out inside the circle, they have more time to
infiltrate small flocks of Reynolds-Vicsek agents and induce a circling
behavior in the final stage.

Finally, we note that \textit{Border 750} is the worst placement strategy,
similar to when it is used in the traveling behaviors.
Also, the \textit{polygon} behavior tends to underperform or match the
performance of \textit{circle}, which tells us that adopting occasional sharper
turns can sometimes be detrimental.

\section{Related Work}
\label{sec:related}
Our work builds upon a series of papers by Genter and Stone examining ways to
use external agents to influence flocking \cite{genter2015placement,
genter2014neighborsorientherd, genter2013visionstationary,
genter2013backsearch, genter2016facegoalfacecurrent,
genter201612steplookahead}.
This prior work studied a number of placement strategies and influencing
agent behaviors, including questions of how best to join or leave a flock in real
scenarios.
Genter also presented results from simulations with
different implementations of Reynold's flocking model, as well as physical
experiments with these algorithms in a small RoboCup setting
\cite{genterthesis}.
This prior work almost exclusively studied small environments, where
density of agents is high, and quick flock formation was virtually guaranteed.
We study two new low-density environments and introduce behaviors to adapt to
the difficulties presented by these new environments.

Jadbadbaie et al. \cite{jad2003convergence} studied Reynolds-Vicsek agents from
an analytical perspective.
Two strong results from this work were that a group of Reynolds-Vicsek
agents in a toroidal setting will eventually converge regardless of initial
conditions, and that in the presence of a single agent with fixed
orientation (analogous to a single influencing agent), all the agents will
converge to that fixed agent's orientation.
This theory provides important context for Genter and Stone's work and the work
that we present here: when the setting is toroidal, convergence is guaranteed,
so the interesting question is how fast we can reach convergence.

Couzin et al. \cite{couzin2005} studied the design of influencing agents
for flocking as well, albeit with a slightly different flocking model.
They proposed an influencing behavior wherein influencing agents adopt
orientations ``in between" their desired goal orientation and the orientations
of their neighbors, in order to still influence their neighbors while not
adopting orientations so extreme that they have no chance of being effective in
the long term.
This is similar in spirit to the motivation behind the multistep algorithm.
We adapted Couzin's algorithm to the new settings, but do not present the
results in this text for space reasons.
The adaptation did not perform as well as the new \textit{multistep} behavior
or the \textit{face} behavior, but it did outperform the \textit{one-step
lookahead} behavior.

Han et al. \cite{han2010teleporting} published a series of papers showing
how to align a group of agents in the same direction.
This work assumed a single influencing agent with infinite speed, and
used this property to construct a behavior that has the influencing agent
fly around and correct the orientation of agents one at a time.
The result is that the Reynolds-Vicsek agents all eventually converge to the
target direction, but are not connected to each other.
In our work, we limit the speed of influencing agents to be the same as the
Reynolds-Vicsek agents to prevent the use of behaviors like this, in hopes
that our results will be more relevant to real applications; we suspect that
influencing agents that act similarly to real agents will be more successful
in real applications.

Su et al. \cite{su2009virtualleaderinformed} studied the question of
flock formation and convergence, but in the context of the
Olfati-Saber flocking model \cite{olfati2006virtualleaderinformed}.
This model assumes the existence of a single virtual leader that non-influencing
agents know about.
The virtual leader plays the role of an influencer, but has special control over
the other agents based on its status.
In our work, we assume that influencing agents do not have any special
interaction rules with Reynolds-Vicsek agents.

Researchers of collective animal behavior have begun using replica conspecifics
in order to influence animal groups across a range of species, from fish to
ducks to cockroaches \cite{Halloy2007, vaughan98, WardFish08}.
Halloy et. al. used robotic influencing agents to control groups of cockroaches;
they exploited the cockroaches' inability to differentiate between real
cockroaches and robotic influencing agents.
Vaugahn et. al. used robotic influencing agents to herd a flock of ducks (on
the ground) to a goal position in a small caged area; their approach used robot
agents to ``push" the ducks from a distance, like a dog herding sheep.

\section{Conclusion}
\label{sec:conclusion}
We have studied the problem of controlling flocks using influencing agents
under two new, more adversarial environments with lower agent density, and
have introduced novel control behaviors for these settings.
In addition to these new algorithms, we have found that in low-density
environments it is more important for influencing agents to
\textit{maintain influence} than it is for them to rapidly turn their
neighbors towards the correct destination.
As a result, earlier results from smaller simulation environments often do
not hold in the environments we introduce.
We found that a multi-stage approach that first embeds influencing
agents in small flocks before attempting to steer these flocks to the goal
direction can be effective in addressing some of these shortcomings.

Although we did not present results from using the multi-stage approach in
the smaller simulation environments from previous work, preliminary
experiments suggest that, in the smaller environments, it is not as effective
as algorithms that optimize for rapid convergence.
Future work could try to find an algorithm that works well in all settings.

It could also explore how to aggregate small flocks into one larger flock.
Many behaviors result in multiple small flocks clustered around
influencing agents that have converged in the sense that they are all facing
the same direction, but remain disconnected from each other.
A successful algorithm would have to change the direction of the flock without
losing individual Reynolds-Vicsek agents on the edges of the flock.

\begin{acks}
The research reported in this paper evolved from a Harvard course project done
by the first two authors and taught by the last two.
We are grateful to Katie Genter for her advice and willingness to share her
flocking code.
\end{acks}

\bibliographystyle{ACM-Reference-Format}
\bibliography{example-bib}

\end{document}